\documentstyle[prd,epsfig,aps]{revtex}

\def\beqra{\begin{eqnarray}}
\def\eeqra{\end{eqnarray}}
\def\beq{\begin{equation}}
\def\eeq{\end{equation}}
\def\ds{\displaystyle}

\def\L{\Lambda}

\def\agt{\stackrel{>}{\sim}}
\def\alt{\stackrel{<}{\sim}}
\begin{document}
\twocolumn

\draft
\input epsf
\twocolumn[\hsize\textwidth\columnwidth\hsize\csname
@twocolumnfalse\endcsname

\title{Dark Energy Condensation}
\author{Massimo Pietroni} 
\address{\it INFN, Sezione di Padova, via Marzolo 8, I-35131, Padova, Italy}
\address{\rm e--mail: pietroni@pd.infn.it}

\maketitle
\begin{abstract}
The two most popular candidates for dark energy, {\it i.e.} a cosmological constant and quintessence, are very difficult to distinguish observationally, mostly because the quintessence field does not have sizable fluctuations. We study a scalar field model for dark energy in which the scalar field is invariant under reflection symmetry $\phi \to -\phi$. Under general assumptions, there is a phase transition at late times ($z\alt 0.5$). Before the phase transition, the field behaves as a cosmological constant. After the phase transition, a time-dependent $\phi$-condensate forms, the field couples with dark matter and develops sizable perturbations tracking those of dark matter. The background cosmological evolution is in agreement with existing observations, but might be clearly distinguished from that of a cosmological constant by future Supernovae surveys. The growth of cosmological perturbations carries the imprint of the phase transition, however a non-linear approach has to be developed in order to study it quantitatively.

\end{abstract}

]
\section{Introduction}
In the past few years, an increasing wealth of data has been shaping a consistent picture of the present state of the Universe: it is spatially flat, mostly made of non-conventional matter --baryons being allowed only up to $\sim 5\%$ of the total energy content-- and accelerating. The physical models accounting for such a picture generally contain two basic ingredients: pressureless dark matter (DM), responsible for the growth of cosmological perturbations via gravitational instability, and negative pressure dark energy (DE),  responsible for the accelerated expansion. The approximate ratio of between DM and DE is around 1:2 today \cite{reviews}. 

The simplest model along these lines is $\Lambda$CDM, in which the role of DE is played by a cosmological constant. It fits very well all the data related with the cosmological background and the perturbations in the linear regime (see for instance \cite{seljak}). Another possibility widely discussed in the literature is quintessence, in which DE has some dynamics, modeled by a scalar field \cite{quinte}. 

In principle, one would like to have clear signatures to distinguish quintessence from $\Lambda$CDM in present or future experiments, but this is not so straightforward.  First of all, existing data already push the present DE equation of state, $w\equiv p/\rho$, very close to the cosmological constant value, $w=-1\pm 0.2$ at $95 \%$ c.l., with at most a very mild evolution up to redshift $z\sim 1$ \cite{seljak}. Secondly, in order to avoid fine-tuning on the initial conditions, the quintessence scalar field is usually taken to be extremely light, with a Compton wavelength corresponding to the present value of the Hubble radius. As a consequence, the scalar field is homogeneous on all observable scales, much like a cosmological constant \cite{quinte}.

Quintessence models suffer from a more theoretical problem, namely the fact that radiative corrections induced by the couplings with the matter fields would generically induce huge corrections to the tree-level mass, thus spoiling the required lightness \cite{light}. So a fine-tuning on the radiative corrections is generally required in these models to keep the scalar field light, besides the one necessary to keep the cosmological constant small.

The purpose of this paper is to try and challenge the general picture of a universe in which two thirds of the universe (the DE) are smoothly distributed, while it is only the remaining third which clumps in structures.

We will discuss a model containing two main ingredients: a scalar field $\phi$ coupled to matter (dark, baryonic, etc.), and a discrete, $Z_2$, symmetry acting upon it, $\phi \to -\phi$. 
Under very general conditions, the scalar field may experience a phase transition. At high matter number density, the $Z_2$ symmetry is restored, the field decoupled, and its fluctuations damped. At low number density the $Z_2$ symmetry is broken by a $\phi$ condensate, $\phi$ and matter couple very efficiently,  and the field fluctuations can grow very large.

In the broken phase, in which we live today, the background energy densities in DE and DM scale with a fixed ratio and both have an effective equation of state in the range $-0.73\alt w \alt -0.5$. Moreover, DE is not smooth on large scales, but has energy density fluctuations tracking the matter number fluctuations. 

So, if one defines DE as the smooth component of the Universe, then in this model --after the phase transition-- there is {\it no} DE. If one defines it as the negative pressure component, then there is {\it only} DE!

There is, however, a difference between the fluctuations in DM and those in DE. While the former can grow and become non-linear, the latter stop growing when the number density  inside the collapsing structures reaches the critical value above which the $Z_2$ symmetry is restored and the coupling between matter and the scalar field vanishes. As we will discuss in sect.~\ref{perturba}, this effect is missed by linear perturbation theory, which very soon becomes unreliable after the phase transition. 

We will work in the framework of scalar-tensor theories of gravity. The choice is motivated by the better status of radiative corrections in this models as compared to minimally coupled quintessence,  but from a cosmological point of view, the coupling between the scalar field and dark matter can be much more generic than that prescribed in this class of theories. The only crucial requirement is a (discrete or continuos) symmetry broken by a $\phi$ condensate.

The action of the model is given by
\beq
S=S_g+S_\phi+S_m\,.
\eeq
$S_g$ is the usual Hilbert action of General Relativity
\[S_g= \frac{M_p^2}{2} \int d^4x \sqrt{-g} \,R\, ,\]
where $M_p=(8 \pi G_N)^{-1/2}$ is the reduced Planck mass,
$S_\phi$ the action for a real scalar field,
\[
S_\phi = \int d^4x \sqrt{-g} \,\left( \frac{1}{2} g^{\mu \nu} \partial_\mu \phi \partial_\nu \phi -V_{tree}(\phi^2)  \right)\, ,
\]
and $S_m$ is the action for all the remaining fields, that is, quark, leptons, DM particles, gauge bosons, GUT particles, etc. which we will call `matter' for simplicity. We assume the following form for $S_m$,
\beq
S_m=  \int d^4x \sqrt{-\tilde{g}}\, {\cal L}_m(\chi_1,\cdots.\chi_N, \tilde{g}_{\mu\nu})\,,
\label{Sm}
\eeq
where the $\chi_i's$ represent all the fields of the model. All the non-gravitational couplings between the scalar $\phi$ and the rest of the world are encoded in the metric 
\beq 
\tilde{g}_{\mu\nu} \equiv  \exp\left(2 b \frac{\phi^2}{\mu^2}\right){g}_{\mu\nu}\,.
\eeq

What we have defined is a scalar-tensor theory (formulated in the Einstein-frame) with a $Z_2$ symmetry imposed on the scalar: $\phi \rightarrow -\phi$.

The paper is organized as follows. In sect.~\ref{radia} we discuss radiative corrections to the scalar field effective action. In sect.~\ref{backgr} we study the background cosmology and in sect.~ \ref{perturba} the growth of cosmological perturbation. Finally, in sect.~\ref{discu}, we summarize our findings and discuss their implications.

\section{Radiative corrections}
\label{radia}

Before discussing the cosmology, let's analyze radiative contributions to the scalar field action\footnote{For a discussion of the case of minimally coupled quintessence, see ref.\cite{doran}}. The contribution from all the loops containing matter fields can be obtained by integrating out the $\chi_i 's$ fields in (\ref{Sm}). The resulting effective action then depends on $\tilde{g}_{\mu\nu}$ only, and its form is dictated by general coordinate invariance as,
\beq S_m \rightarrow \int d^4x \sqrt{-\tilde{g}}\, \left[ \Lambda_0^4  + \Lambda_1^2  \tilde{R} +\cdots \right]\,,
\eeq
where $\Lambda_0$ and $\Lambda_1$ are ultraviolet (UV) cutoff, $\tilde{R}$ is the Ricci scalar built out of the metric $\tilde{g}_{\mu\nu}$ and the ellipses stand for subdominant logarithmically divergent and finite terms. 
In terms of the metric $g_{\mu\nu}$ we have,
\beqra
&& \ds 
\int d^4x \sqrt{-g}\, \left[ e^{4 b \phi^2/\mu^2} \Lambda_0^4 + \right. \nonumber \\
&& \ds \left. e^{2 b \phi^2/ \mu^2} \Lambda_1^2 \left(R  -\frac{6b}{\mu^2} \Box \phi^2 -\frac{24 b^2 \phi^2}{ \mu^4}  g^{\mu\nu} \partial_\mu \phi \partial_\nu \phi \right) +\cdots \right] \,,\nonumber \\
&&
\label{effe}
\eeqra

Taking the $\phi$ field on its expectation value, the first term contributes to the cosmological constant, and has to be fine-tuned to $\L_0 \alt 10^{-3} \,{\mathrm eV}$. Derivatives of this term with respect to $\phi$ give corrections to the tree-level terms in $V_{tree}(\phi)$. In particular, we will focus on the corrections to the tree-level mass term which, as we will see (see eq.~(\ref{Vtree})) is $O((10^3 {\rm eV})^4/\mu^2)$. The correction proportional to $\Lambda_0^4/\mu^2$ is then subdominant as a consequence of the same fine-tuning mentioned above.

 The terms proportional to $\Lambda_1^2$ include corrections to the tree-level kinetic terms. We require $\Lambda_1\alt \mu$ in order to keep them subdominant. The $\Lambda_1^2 R\sim \Lambda_1^2 H^2 $ term gives also contributions to the mass of the scalar field, which is however negligible if $\Lambda_1 \alt M_p$. 
 
Summarizing, general coordinate invariance ensures that, once a single fine-tuning is done on the cosmological constant, all the radiative corrections coming from the matter sector do not destabilize the lightness of the scalar field. In other words, it forbids a $\Lambda_2^2 \phi^2$ term in eq.~(\ref{effe}), with the UV cutoff $\Lambda_2$ independent on $\Lambda_0$.
If, for instance, we take $\mu=O(M_p)$ as in usual quintessence models, the corrections to the mass of the scalar field today are smaller than $H_0^2 \simeq 10^{-33}  {\rm eV}$, without any further fine-tuning. For comparison, in minimally coupled quintessence models the lightness of the scalar field generally requires a further fine-tuning besides that for the cosmological constant \cite{light}, or a new symmetry principle, as in \cite{barb}.

Scalar field self-interactions coming both from the tree-level action and the effective one, eq.~(\ref{effe}), can be shown not to destabilize the tree-level mass provided the UV cutoff is $\alt \mu$. Taking $\mu$ larger than, say, a TeV, this implies no fine-tuning on the scalar-sector. 

Finally, graviton loops have to be considered. Calling $\Lambda_g$ the UV cutoff for this kind of contributions, one can see that the most dangerous terms involve couplings  from eq.~(\ref{effe}), of the form
\beq 
e^{2 b \phi^2/\mu^2} \Lambda_1^2 \frac{(\partial h)^2}{M_p^2}\,,
\label{gravi}
\eeq
where $h$ is the metric fluctuation, $g_{\mu\nu} = \eta_{\mu\nu}+h_{\mu\nu}/M_p$. 
These terms give a quartically divergent contribution both to the cosmological constant and to the scalar field mass. Requiring that the correction to the cosmological constant is smaller than $O(10^{-3} {\mathrm eV})$, one obtains the bound  $(\Lambda_1/M_p)^2 (\Lambda_g/10^{-3} {\rm eV})^4 \alt 1$, which, again, ensures that the mass correction is subdominant.

\section{Background evolution}
\label{backgr}
In the Einstein frame the field equations have a simple form. The Friedmann equation is the usual one for a flat universe with radiation, matter, and a canonical scalar field, provided the energy densities of radiation and matter satisfy the modified Bianchi identity
\[
 d(\rho a^3) + p da^3=(\rho-3p)a^3 \, d\left(\frac{b \phi^2}{\mu^2}\right)\,,
\]
from which we read that matter scales as $\rho_m \sim \exp(b \phi^2/\mu^2) a^{-3}$, while radiation has the usual behavior $\rho_{rad}\sim a^{-4}$.

The scalar field dynamics is governed by the equation
\beq
\ddot{\phi} + 3 H \dot{\phi} = -\frac{dV_{tree}}{d \phi} - \frac{2 b \phi}{\mu^2} \rho_b (1-3w_b)\,
\label{feq}
\eeq
where $\rho_b\equiv \rho_{rad}+ \rho_m$ and $w_b\equiv p_b/\rho_b$. 

At early times, when $\rho_b \gg V_{tree}$, only the second term in the RHS matters. This would have the effect of driving the field towards the symmetric point $\phi=0$, but  is proportional to the trace of the energy momentum tensor, which vanishes for a Universe dominated by a conformally invariant gas of non-interacting relativistic particles. However, in the real world, conformal invariance is broken even during radiation domination, by two effects. 

The first is the presence of mass thresholds \cite{dampol}. Each time a particle in  thermal equilibrium becomes non-relativistic, it gives a non-vanishing contribution to the trace of the energy momentum tensor,
\begin{equation}
\frac{\rho_A- 3 p_A}{\rho_{rad}}\simeq\frac{15}{\pi^4}\frac{g_A}{g_\star} y_A^2  F[y_A]\,,
\end{equation}
with $y_A\equiv m_A/T$, $g_A$ the number of degrees of freedom of the particle A, $g_\star$ the number of relativistic degrees of freedom and
\begin{equation}
F[y_A] \equiv
\int_0^\infty dx \frac{x^2}{\varepsilon_A [\exp(\varepsilon_A) \pm 1]}\,,
\label{Ffm}
\end{equation}
where $\varepsilon_A\equiv (y_A^2+x^2)^{1/2}$ and the minus (plus)
sign in the denominator of the integrand holds for bosons (fermions).
The function $F[y_A]$ is $O(1)$ around $T=m_A$, quadratically suppressed at high temperatures and  exponentially suppressed at low ones.

The second source of breaking of conformal invariance is given by radiative corrections which, for a plasma of the $SU(N_c)$ gauge theory with coupling $g$ and $N_f$ flavors, give the following {\it trace anomaly} \cite{trace}
\[ \left. 1-3w_b\right|_{\rm t.\,a.} = \frac{5}{6 \pi^2} \frac{g^4}{(4\pi)^2}\frac{(N_c+\frac{5}{4}N_f)(\frac{11}{3}N_c-\frac{2}{3}N_f)}{2+\frac{7}{2}\left[N_cN_f/(N_c^2-1)\right]}\,,
\]
leading to a constant contribution which, for typical gauge theories, can be of order $1-3w_b \sim 10^{-2}-10^{-1}$. At high temperatures, this effect dominates over that of mass thresholds, which is suppressed by the number of relativistic degrees of freedom, $g_\ast$. 

The joint effect of these two contributions was studied for instance in ref. \cite{catena}, where it was shown that the field can have a sizable evolution during radiation domination, reaching phenomenologically acceptable values well before nucleosynthesis. In the present case, the field will be driven to the symmetric point $\phi=0$ with an efficiency increased with respect to ref.\cite{catena} by the fact that we will take  $\mu$ much lower than the Planck mass. This provides a post-equivalence initial condition which is completely independent on the conditions of the field at a early time, e.g., after inflation.

We now concentrate on the epoch of matter domination and on the contributions to (\ref{Sm}) from non relativistic cosmological relics, in particular DM particles, similar contributions (though subdominant) arising also from baryons and neutrinos. The action for a single, non-relativistic, particle of mass $m$ is
\beq
- m \int d\tilde{s} = - m  \int e^{b \phi^2/ \mu^2} ds\,,
\eeq
which  corresponds to a field dependent mass  $m e^{b \phi^2/\mu^2}$. 

To facilitate the discussion we will absorb the second term in the RHS of eq.~(\ref{feq})  into the effective potential for $\phi$,
\beq
V(\phi) =V_{tree}(\phi^2) + m e^{b \phi^2/ \mu^2} n +\ldots\,
\label{Vphi}
\eeq
where the average number density is defined as $n= \sum_i m_i n_i / m$, with $m =\sum_i m_i$, and the sum runs over all the  non relativistic particles.  The ellipses represent the radiative contributions, which, based on the discussion in the previous session, we assume to be subdominant.  

If the curvature of the effective potential (\ref{Vphi}) is always positive at the symmetric point $\phi =0$, then the scalar field is always fixed at that point, its coupling with matter, $2 b \phi/\mu^2$ is zero, and the cosmology is that of a $\Lambda$CDM model.

We will instead consider the case in which the tree level effective potential has negative curvature in $\phi=0$, while $b>0$. For definitness, we will use the potential
\beq
 V_{tree}(\phi^2) =V_0 \, e^{-\phi^2/\mu^2} \,,
 \label{Vtree}
 \eeq
however, using other forms will not change our main results considerably, since, as we will see, during all the cosmological evolution the field samples only a limited range of values, $|\phi/\mu| \alt 1$, so that the two parameters $V_0$ and $\mu$ could be effectively traded with the value of a generic, $Z_2$-symmetric, potential at the origin and its second derivative there.
 
The immediate consequence of our choice for $V_{tree}$ and $b$ is a number density driven phase transition. At high number density $n$,  $\phi$ vanishes at the minimum of (\ref{Vphi}). In this regime the relic particles' masses are constant and the two contributions in (\ref{Vphi}) scale as cosmological constant and matter, respectively. As the universe expands, if the number density drops below the critical value
\beq
\bar{n} \equiv \frac{V_0}{b m}\,,
\label{nbar}
\eeq
the effective potential is minimized by 
\beq
\frac{\phi^2}{\mu^2} = \frac{1}{b+1}\log\left(\frac{\bar{n}}{n}\right) \,\,\,\,\quad\quad\qquad(
\,\,n\le\bar{n})\,.
\eeq

Thus, on the minimum, we have the two regimes for the effective potential,
\beqra
\langle V \rangle = 
&& V_0 \left(1 + \frac{1}{b}\frac{n}{\bar{n}} \right) \,\,\,\,\,\,\qquad\quad\quad(n\ge \bar{n}),\\
&&\nonumber\\
&&V_0 \frac{b+1}{b} \left(\frac{n}{\bar{n}}\right)^{\frac{1}{b+1}}\,\,\,\,\,\,\qquad\quad(n\le \bar{n}),
\label{minima}
\eeqra
where, as we anticipated in the Introduction, after the phase transition DM and DE get diluted at the same rate. 
Notice that $\langle V \rangle$ depends on three parameters ($V_0$, $\bar{n}$, and $b$). The $\mu$ parameter disappears from the background energy density if the field tracks the minimum and the kinetic energy contribution is negligible, which happens for $\mu \alt 10^{-2} M_p$, as we will see.

To study this parameter space (which, lacking a theoretical motivation for a particular value for $b$, contains one more parameter than $\Lambda$CDM) we proceed as follows. First, by going to the high density regime  of (\ref{minima}), we require that the matter density at recombination agrees with the result from WMAP \cite{wmap}. Trading the critical density $\bar{n}$ with the redshift of the phase transition, $\bar{z}$,
$(1+\bar{z}) \equiv \left(\bar{n}/{n_0}\right)^{1/3}$, we obtain
\beq (1+\bar{z})=\left[(b+1) \Omega_M^0\right]^{-(b+1)/3b}\,,
\label{zbar}
\eeq
which fixes $\bar{z}$ ($\bar{n}$) as a function of $b$ and the measured $\Omega_M^0 = 0.27 \pm 0.04$ \cite{wmap}.
Armed with this relation, we can compute two observables: the CMB shift parameter \cite{shift},
\beq 
{\cal R} = \left[(b+1) (1+\bar{z})^{3b/(b+1)} \right]^{-1/2} \int_0^{z_{dec}} dz' \frac{H_0}{H(z')}\,,
\eeq
which is measured to be ${\cal R}=1.716\pm 0.062$ by WMAP  \cite{wmap}, and the luminosity distance -redshift relation measured using type Ia supernovae (SNeIa) \cite{SNe,sngold},
\beq
H_0 d_L(z) = (1+z)\int_0^z dz' \frac{H_0}{H(z')}\,.
\label{dlz}
\eeq
Notice that both these observables are independent on the absolute scale of $H_0$, that is,  they do not depend on the parameter $V_0$. Then, once (\ref{zbar}) is used, the CMB shift parameter and the SneIa luminosity distance measurements depend on the parameter $b$ only.

The measurements of ${\cal R}$ constrain $b$ to be $\agt 0.5$, corresponding to $\bar{z}\alt1.47$. On the other hand, the requirement that the phase transition takes place before today, gives the upper bound on $b$, $b\le 1/\Omega_M^0-1=2.7$.

\begin{figure*}[t]
\epsfxsize=3. in 
\epsfbox{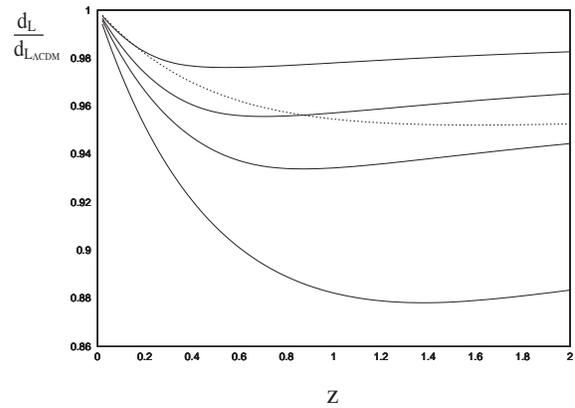}
\caption{Ratios of the luminosity distance to that of a reference $\Lambda$CDM model with $\Omega_M =0.27$. The continuos lines correspond, from bottom to top, to $b=0.5,\,0.8,\,1,\,$ and $1.3$. The dotted line corresponds to a flat model with $ \Omega_M=0.27$ and the rest in a fluid with constant equaion of state, $w=-0.8$. }
\label{ratio}
\end{figure*}

In Fig. ~\ref{ratio} we plot the luminosity distance {\it vs} redshift, normalized to a fiducial flat $\Lambda$CDM model with $\Omega_M^0=0.27$, for different values of $b$. The value of $\bar{z}$ has been fixed in order to have the same  $\Omega_M^0$, according to eq.~(\ref{zbar}). We also plot, with the dotted line, the result for a quintessence model with constant equation of state $w=-0.8$.

The SNeIa data turn out to give a stronger constraint than ${\cal R}$. 
We tested the distance-redshift relations of our models against the ``gold" set of 157 SNeIa of ref. \cite{sngold}, using flux-averaging statystics as implemented in \cite{wang}.
We find $b \ge 1$ (corresponding to $\bar{z} \le 0.51$) at $95 \%$ c.l. \footnote{I thank Luca Scarabello for providing help on this part of the analysis.}

Summarizing, the background and CMB measurements are compatible with a late time ($\bar{z}\alt 0.5$)  phase transition from a high density phase in which the dynamics is that of a $\Lambda$CDM model to a low density one in which the background evolves according to  eq.~(\ref{minima}). In this scenario, the energy density of the Universe today is dominated by a DM-DE fluid in which the two components scale at a fixed ratio $\Omega_M/\Omega_V=b$ and are diluted as $(1+z)^{3/(b+1)}$, that is with an effective equation of state $w=-b/(b+1)$. The bounds on $b$ discussed above correspond to $ -0.73\alt w \alt -1/2$ today. For comparison, assuming a constant $w$, the SneIa limit is $w< -0.78$ at  $95 \%$ c.l. \cite{sngold}.

\begin{figure*}[t]
\epsfxsize=3. in 
\epsfbox{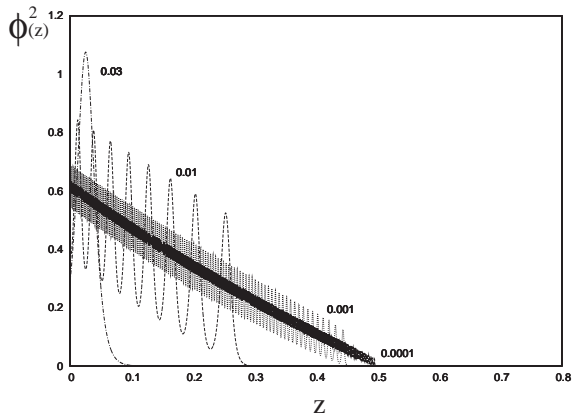}
\caption{Evolution of the square of the scalar field for different values of the $\mu/M_p$ ratio (for $\mu/M_p \agt 0.05 $ the field does not move sensibly). We set $b=1$, the lowest value allowed by SnIa data, which corresponds to $\bar{z}=0.51$.}
\label{field}
\end{figure*}

The analysis above assumed  that after the phase transition the field adiabatically follows the moving minimum. In Fig.~\ref{field} we plot  the field evolution as resulting from the equation of motion for different values of $\mu/M_p$. For large values of $\mu$ there is basically no evolution of the field expectation value before today. So, the only feature of the model in this case is an extremely flat potential, giving a decoupled scalar field in the spectrum with mass $\alt H_0 \sim (10^{-33} {\mathrm eV})$. As $\mu$ is decreased, say, for $\mu/M_p \alt 0.01$, the field follows the minimum while kinetic terms do not contribute to the energy density sensibly. Notice the oscillations around the moving minimum. As $\mu$ decreases, the frequency of the oscillation increases due to the increased scalar field mass,
\beqra
m^2_\phi  = \langle V''(\phi) \rangle && = \frac{2 V_0}{\mu^2}\left(\frac{n}{\bar{n}}-1\right)\,,\quad\quad\quad \quad\quad\quad\quad(n\ge \bar{n})\nonumber\\
&& =  \frac{4 V_0}{\mu^2} \left(\frac{n}{\bar{n}}\right)^{1/(b+1)} \log \frac{\bar{n}}{n}\,,\,\quad\quad\quad (n\le \bar{n})\nonumber\\
&&
\label{mass}
\eeqra
which is $M_p^2/\mu^2$ times $O(H^2)$. On the other hand, the amplitude decreases, since the field starts moving earlier after the phase transition. 

 We have verified that, for $\mu/M_p \alt 0.01$, the field  oscillations leave no observable imprint in the luminosity distance-redshift relation, since they get averaged out in the integral of eq.~(\ref{dlz}). 

\section{Cosmological Perturbations}
\label{perturba}
More dramatic signatures, potentially such as to allow a clear distinction from  $\Lambda$CDM, may come from perturbations. In particular, the phase transition turns on a coupling between the scalar field and matter which induces large perturbations of $\phi$.

Linear perturbations in DE-DM coupled systems have been already discussed in the literature \cite{coinc,amquer,MBP,CPS}. Non linear approaches have been attempted in \cite{MPS,Bonometto,gubpeeb}.

As compared to the uncoupled case (normal quintessence scalar fields) two main additional effects have to be taken into account: 

1) particles' trajectories may deviate from geodesics:

\beq
\dot{\theta} +\left( \frac{\dot{a}}{a}+2 b \frac{\phi\dot{\phi}}{\mu^2}\right)\theta= 2 b k^2\frac{\phi \delta\varphi}{\mu^2}\,,
\label{theta}
\eeq
where $\delta \varphi$ is the field fluctuation, $\theta=i k^j v_j$ is the divergence of the comoving velocity, and we work in synchronous gauge, see for instance \cite{MB}.
Notice that, as a consequence of the $Z_2$-symmetry the field is decoupled from matter in $\phi=0$, so the particles follow geodesics in the high-density phase, that is, for $z>\bar{z}$ and inside objects with number density $n>\bar{n}$. For comparison, in the models considered in ref.~\cite{coinc}, the RHS is always active, which gives strong constraints on the strength of the coupling;

2) the field's fluctuations have a new source term on the RHS:
\beqra
&& \delta\ddot{\varphi}+2 \frac{\dot{a}}{a}\delta\dot{\varphi}+(k^2+ a^2 m_\phi^2)\delta\varphi= 
\nonumber \\
&& \quad\quad\quad -\frac{1}{2} \dot{h}\dot{\phi} -2 b \frac{\phi}{\mu^2}  e^{b \phi^2/\mu^2} m \,\delta n \, a^2 \,,
\eeqra
with $m^2_\phi$ given in eq.~(\ref{mass}). The last term in the RHS induces potentially large field fluctuations after the phase transition.
\beq
\frac{\delta\varphi}{\phi} \simeq -2\frac{V_0}{\mu^2 (k^2/a^2 +m_\phi^2)}\left(\frac{n}{\bar{n}}\right)^{1/(b+1)}  \frac{\delta n}{n}\,.
\label{fluctuations}
\eeq
For subhorizon scales with  $H< k/a < m_\phi$ these are of the same order as $\delta{n}/n$,
\beq
\frac{\delta\varphi}{\phi} \simeq - \left( 2 \log \frac{\bar{n}}{n}\right)^{-1} \frac{\delta n}{n}\,.
\eeq
Notice the minus sign, implying that the DM particles' mass is smaller in overdense regions.
The divergence as $n\to \bar{n}$ is cured by the $k^2/a^2$ term in (\ref{fluctuations}), and is a general manifestation of second order phase transitions, that is, the presence of large fluctuations on large scales close to the transition point.

\begin{figure*}[t]
\begin{center}
\includegraphics[width=3. in]{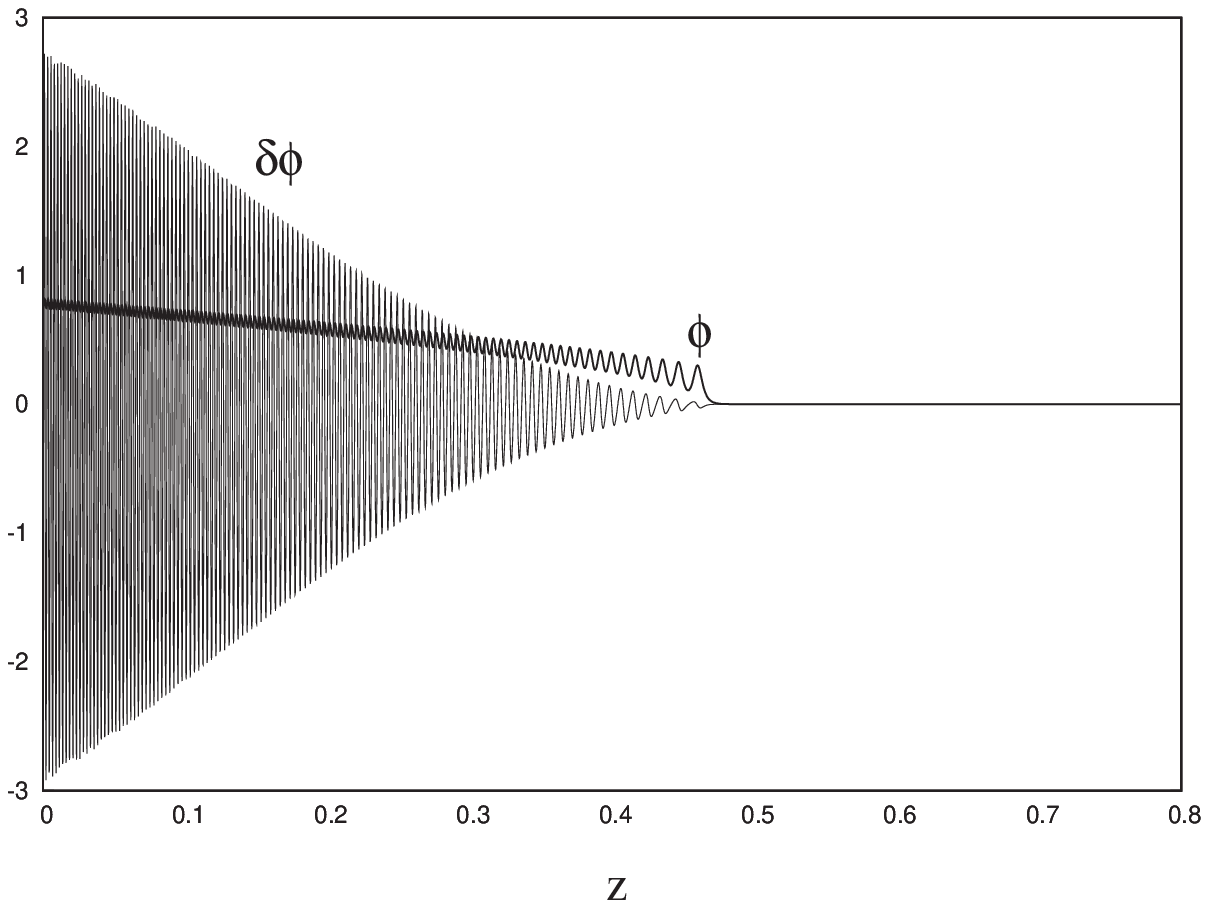}
\includegraphics[width=3. in]{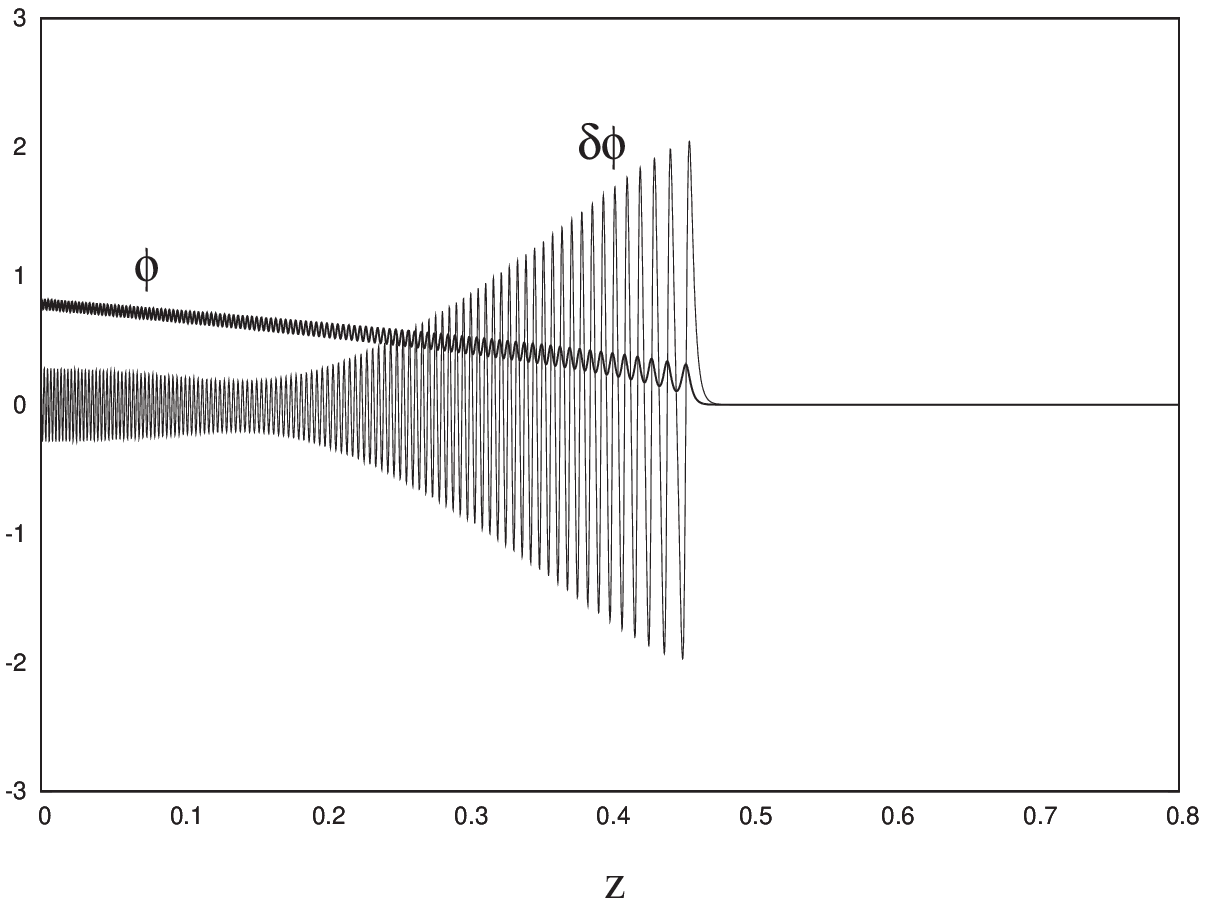}
\caption{Evolution of the field and its fluctuation in linear perturbation theory (upper panel) and after substituting $n$ with $n(1+\delta_n)$ in the effective potential (lower panel). We set $\mu/M_p=0.001$, $b=1$, $k/aH=1$.}
\label{fluct}
\end{center}
\end{figure*}

\begin{figure*}[t]
\epsfxsize=3. in 
\epsfbox{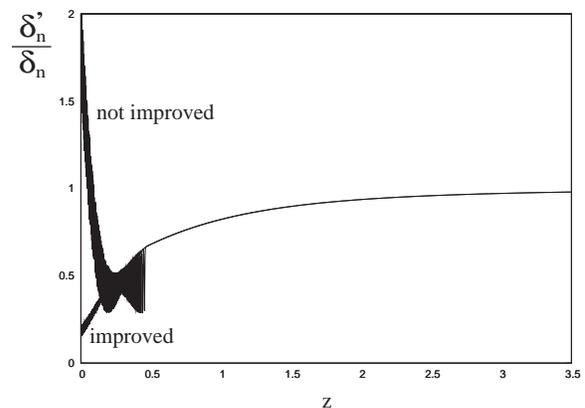}
\caption{The ratio $\delta_n^\prime/\delta_n$ before and after the improvement. Same parameters as in Fig.~\ref{fluct}.}
\label{dnp}
\end{figure*}

In Fig.~\ref{fluct} we plot the field average and fluctuation in linear perturbation theory (upper panel). After the phase transition, fluctuations grow until they become larger than the field  average and perturbation theory is no longer reliable. Correspondingly, the fluctuation in number density, $\delta_n\equiv \delta n/n$ blows up at late time (Fig.~\ref{dnp}, curve `Not improved', where we plot $d \log \delta_n/d \log a$). The behavior of the perturbations seems then very different than for $\Lambda$CDM and in strong disagreement with the result extracted from 2dF survey, from which one extracts \cite{2df,verde} 
\beq
d \log \delta_n/d \log a =0.51\pm 0.11,
\eeq
at the effective redshift of the survey, $z\sim 0.15$.

However, linear perturbation theory misses a crucial physical effect, which drastically affects the results. Indeed, when a number overdensity grows beyond the critical value $\bar{n}$, the $Z_2$ symmetry gets restored and matter and the scalar field decouple again. As a consequence, above the critical value, there is no contribution from the scalar force and the overdensity  grows under the standard effect of gravity alone. On the other hand, in linear perturbation theory, the strength of the DM-scalar coupling is always proportional to the background value of $\phi$, which is non-vanishing for $z\le \bar{z}$, independently of the value of the overdensity.

As a first step towards a more realistic description of the growth of perturbations in this model, we `improved' the equation of motion for the average field by substituting $n$ with $n(1+\delta_n)$ in the effective potential of eq.~(\ref{Vphi}). The results are plotted in the lower panel of Fig.~\ref{fluct} and on Fig.~\ref{dnp} (`Improved'). As we see, at late times the field fluctuations are smaller than the  average, and the growth of density perturbations is damped much like in $\Lambda$CDM, signaling a better behavior of the improved perturbation theory. However, considering smaller length scales, the effect of the scalar force  becomes more and more important (see the RHS of eq.~(\ref{theta})) and the improved perturbation theory fails as well.

\begin{figure*}[t]
\epsfxsize=3. in 
\epsfbox{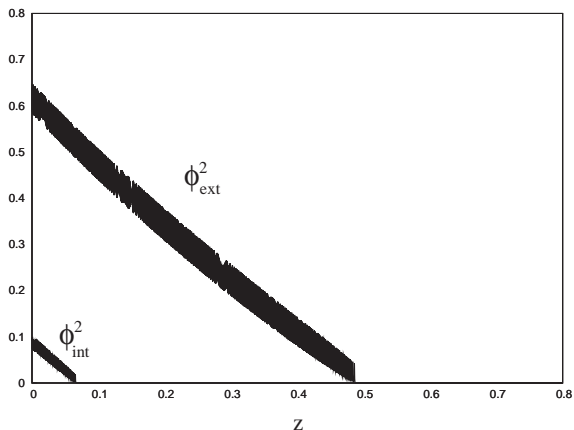}
\caption{The square of the scalar field inside and outside a collapsing spherical overdensity for $\mu/M_p=0.001$, $b=1$.}
\label{sp_field}
\end{figure*}

\begin{figure*}[t]
\epsfxsize=3. in 
\epsfbox{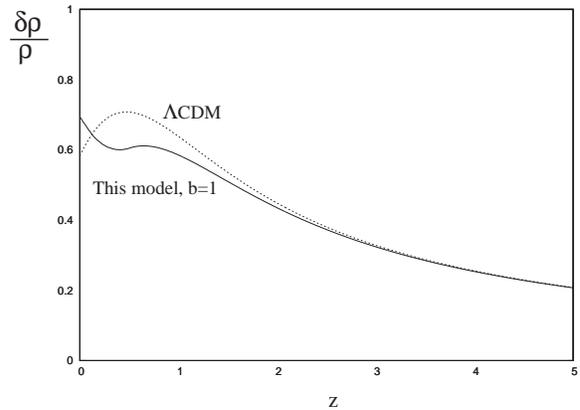}
\caption{Evolution of  $\delta\rho/\rho = \Omega_M  \delta\rho_M/\rho_M+\Omega_V  \delta\rho_V/\rho_V$  for the overdensity in Fig.~\ref{sp_field}. For comparison we plot also the evolution of an overdensity with the same initial condition in the $\Lambda$CDM model.}
\label{sp_dndr}
\end{figure*}

Further insight on the growth of perturbations in this model can be gained in a spherical collapse model. If the mass of the scalar is much larger than the inverse radius of the collapsing sphere, $m_\phi \gg R^{-1}$, that is, using eq.~(\ref{mass}), for $\mu/M_p\ll H R$,  the field  inside the sphere deviates from the background value. In Fig.~\ref{sp_field} we plot the evolution of the field (squared) inside and outside a collapsing spherical overdensity, and in Fig.~\ref{sp_dndr} the corresponding growth of the energy overdensity, $\delta \rho/\rho = \Omega_M  \delta\rho_M/\rho_M+\Omega_V  \delta\rho_V/\rho_V$, where $\rho_M= m \exp(b \phi^2/\mu^2) n$ and $\rho_V = V_{tree}(\phi^2)$. In this example, the phase transition inside the sphere takes place much later than in the background. For slightly larger  overdensities today, but still in the range $\alt 1$, the phase transition inside does not take place at all before today. Such a collapsing sphere is always blind to the scalar force. While individual particles feel a strong effect from the scalar force, moderately overdense structures collapse and attract each other as if there were only gravity. 

For larger values of $b$ the background phase transition takes place later and the behavior is more and more similar to that of standard $\Lambda$CDM.

Inside our Galaxy, the number density today is $\gg \bar{n}$. Then, the scalar field is decoupled from matter and is not detectable via solar system -- or laboratory-- tests of General Relativity. For the same reason, the Newton constant and particle's masses are not varying now and have the same values they had at high redshifts $z \ge \bar{z}$, in particular during nucleosynthesis and matter-radiation decoupling. In this respect, the scalar field behaves similarly to the `chameleon' field introduced in ref.\cite{chameleon}.

On the other hand, individual matter particles fluctuating outside the overdense regions, have time-dependent-- and increasing-- masses today. 

\section{Discussion}
\label{discu}

Models in which DE and DM have extra --non-gravitational -- interactions have been already investigated in the literature (see, for instance, \cite{andcar,MBP,Bartolo,coinc,MPS,Bonometto,chaply}).  The present model differs from these  works in two respects. First of all, in our model, the extra-interaction has been silent for most of the cosmological history, allowing a close to standard background evolution and growth of perturbation. By contrast, in the previous literature the coupling was assumed to be always effective (although evolving, as in \cite{Bartolo}). As a consequence, strong bounds on it were imposed by CMB and Large Scale Structure (\cite{coinc,amquer}). 

The second difference is in the scalar field mass, and then on the scale of perturbations. The requirement of being on an attractor solution implies a  Compton wavelength of order $H^{-1}$ , and then basically no scalar field perturbations on sub-horizon scales, much like in the case of uncoupled quintessence \cite{quinte}.  On the other hand, in this paper the Compton wavelength of the scalar field is $O(\mu/M_p) H^{-1}$, where $\mu$ can take values much lower than $M_p$ (actually, as we see from Fig.~\ref{field}, the lower $\mu$ the better the field follows the moving minimum). The independence of the late-time cosmology on the initial conditions is not achieved in this model by an attractor, but by symmetry. The field is on the symmetric point $\phi=0$ until the phase transition, and then tracks the moving minimum.

A generic expectation from  a phase transition of the type described in this paper is a modification in the growth of cosmological perturbations at low redshifts, likely in the form of an oscillatory behavior of the growth exponent, see figs.~\ref{dnp}, \ref{sp_dndr}. However, in order to extract reliable predictions, more work has to be done on a non-linear approach incorporating the physical effect of symmetry restoration inside high-density regions.

\section{Acknowledgment}
We would like to thank Denis Comelli for discussions and  Luca Scarabello for providing help in enforcing the constraints from SNeIa data.

\end{document}